\documentclass[pra,twocolumn,amsmath,amssymb,nofootinbib]{revtex4}

\usepackage{graphicx}

\begin{document}
\title{Cryptographic quantum bound on nonlocality}
\author{Satoshi Ishizaka}
\affiliation{Graduate School of Integrated Arts and Sciences,
Hiroshima University,
1-7-1 Kagamiyama, Higashi-Hiroshima, 739-8521, Japan}
\date{\today}
%%%%%%%%%%%%%%%%%%%%%%%%%%%%%%%%%%%%%%%%%%%%%%%%%%%%%%%%%%%%%%%%%%%%%%%%%%%%%%%
%
%%%%%%%%%%%%%%%%%%%%%%%%%%%%%%%%%%%%%%%%%%%%%%%%%%%%%%%%%%%%%%%%%%%%%%%%%%%%%%%

\begin{abstract}
Information causality states that the information obtainable by a receiver
cannot be greater than the communication bits from a sender, even if they
utilize no-signaling resources. This physical principle successfully explains
some boundaries between quantum and postquantum nonlocal correlations, where
the obtainable information reaches the maximum limit. We show that
no-signaling resources of pure partially entangled states produce randomness
(or noise) in the communication bits, and achievement of the maximum limit is
impossible, i.e., the information causality principle is insufficient for the
full identification
of the quantum boundaries already for bipartite settings. The nonlocality
inequalities such as so-called the Tsirelson inequality are extended to show
how such randomness affects the strength of nonlocal correlations. As a result,
a relation followed by most of quantum correlations in the simplest Bell
scenario is revealed. The extended inequalities reflect the cryptographic
principle such that a completely scrambled message cannot carry information.
\end{abstract}

%%%%%%%%%%%%%%%%%%%%%%%%%%%%%%%%%%%%%%%%%%%%%%%%%%%%%%%%%%%%%%%%%%%%%%%%%%%%%%%
%
%%%%%%%%%%%%%%%%%%%%%%%%%%%%%%%%%%%%%%%%%%%%%%%%%%%%%%%%%%%%%%%%%%%%%%%%%%%%%%%
\maketitle
%%%%%%%%%%%%%%%%%%%%%%%%%%%%%%%%%%%%%%%%%%%%%%%%%%%%%%%%%%%%%%%%%%%%%%%%%%%%%%%
%
%%%%%%%%%%%%%%%%%%%%%%%%%%%%%%%%%%%%%%%%%%%%%%%%%%%%%%%%%%%%%%%%%%%%%%%%%%%%%%%
\section{Introduction}
\label{sec: Introduction}

It was shown by Bell that the nonlocal correlations predicted by quantum
mechanics are inconsistent with local realism \cite{Bell64a}. The nonlocal
correlations do not contradict the no-signaling principle that prohibits
instantaneous communication. However, it was found that the set of quantum
correlations is strictly smaller than the set of no-signaling
correlations \cite{Tsirelson80a,Popescu94a}. Concretely, a particular type of
the Bell inequality, the Clauser-Horne-Shimony-Holt (CHSH)
inequality \cite{Clauser69a}, was shown to be violated up to $4$ in general
no-signaling correlations \cite{Popescu94a}, while from the Tsirelson
inequality \cite{Tsirelson80a} the violation is bounded by $2\sqrt{2}$ in
quantum correlations. Since then, many efforts have been made to search for a
simple physical principle to close this discrepancy.
See \cite{Brunner14a,Popescu14a,Oas16a} for a good review.

Information causality (IC) \cite{Pawlowski09a} is such a physical principle.
Consider two remote parties, Alice and Bob, who share no-signaling nonlocal
resources such as entangled states. When Alice sends a message to Bob, IC
states that
the total information obtainable by Bob cannot be greater than the number of
the message bits even if they utilize the no-signaling
resources \cite{Pawlowski09a}. A powerful necessary condition for respecting
the IC principle was derived by considering an explicit communication
protocol \cite{Pawlowski09a}. The condition, called the IC inequality
hereafter, successfully explains the Tsirelson inequality, and even explains
some curved boundaries between quantum and postquantum
correlations \cite{Pawlowski09a,Allcock09a}. At those quantum boundaries, the
protocol achieves the maximum limit of the obtainable information (the number
of the message bits). It is then expected that, for every quantum boundary,
there exists a protocol for which the maximum limit is achieved.

Apart from searching for physical principles, the identification of the
quantum boundaries is originally a difficult problem. Indeed, the analytical
necessary and sufficient criterion for the identification has not been given
yet even in the simplest Bell scenario, although the Tsirelson-Landau-Masanes
(TLM) criterion \cite{Tsirelson87a,Landau88a,Masanes03a} is known for a case of
unbiased marginal probabilities (explain later).

In this paper, we show that pure partially entangled states, which were shown
to give rise to boundary
correlations \cite{Liang11a,Acin12a,Ramanathan15a,Zhen16a}, produce randomness
in the message, and achievement of the maximum limit is
impossible no matter what protocol is executed. Hence, the IC principle is
insufficient for the full identification of the quantum boundaries already for
bipartite settings (similar results have been obtained in multipartite
settings \cite{Gallego11a,Yang12a}). We extend the nonlocality inequalities to
include the effects of the randomness. As a result, a relation followed by most
of quantum correlations, including both cases of unbiased and biased marginals,
is revealed. Moreover, we show that the derived inequalities reflect the
cryptographic principle such that a completely scrambled message
cannot carry information \cite{Shannon49a}. The inequalities reflecting the
cryptographic principle contain a quantity defined in quantum mechanics, and
the principle cannot immediately exclude postquantum correlations by itself,
but tells us a way to determine the quantum boundaries. Note that a
similar principle for a different type of randomness was considered
in \cite{Carmi15a}.

Let us here show a simple example to clarify what we mean by randomness.
Consider the simplest Bell scenario, where Alice is given a random bit $x$ and
performs the measurement on the partially entangled state
$\sqrt{2/3}|00\rangle\!+\!\sqrt{1/3}|11\rangle$
in the basis $|0\rangle/|1\rangle$ for $x\!=\!0$
or $(|0\rangle\pm|1\rangle)/\sqrt{2}$ for $x\!=\!1$,
and obtains the outcome $a$.
Suppose that Bob somehow knows that $x$ was 1.
However, he still cannot determine the value of $a$ completely, because his
local state is $\sqrt{2/3}|0\rangle\!+\!\sqrt{1/3}|1\rangle$ for $a\!=\!0$
or $\sqrt{2/3}|0\rangle\!-\!\sqrt{1/3}|1\rangle$ for $a\!=\!1$, which
are nonorthogonal. This means that $a$ has some uncertainty indeterminable for
Bob, and the uncertainty acts as randomness (or noise)
if $a$ is used for the transmission of the information of $x$
(the details are discussed in
Sec.\ \ref{sec: Information theoretical aspects}). As a result, the obtainable
information of Bob is reduced. This affects the quantum bound of the CHSH
inequality, and the violation is reduced to $2\sqrt{17/9}$ [see
Eq.\ (\ref{eq: weighted Tsirelson bound}) below] from the maximum value of
$2\sqrt{2}$.

This paper is organized as follows.
In Sec.\ \ref{sec: Bounds on nonlocality}, we extend the Tsirelson
inequality, the IC inequality, and the Landau inequality (of the TLM criterion)
by including the state-dependent quantity measuring the orthogonality between
Bob's local states, because nonorthogonality is key to the randomness as in
the above example.
In Sec.\ \ref{sec: Tightness of bounds}, we discuss the tightness of
the derived Landau-type inequality, and show that the inequality is widely
saturated for both boundary and non-boundary correlations even in the case of
biased marginals. We then discuss the information theoretical aspects of the
derived IC-type inequality in
Sec.\ \ref{sec: Information theoretical aspects}, where the connection to the
cryptographic principle and the insufficiency of the IC principle are shown.
In Sec.\ \ref{sec: Boundary condition}, we show an example of how to determine
the quantum boundaries under the cryptographic principle. A summary is given in
Sec.\ \ref{sec: Summary}.

%%%%%%%%%%%%%%%%%%%%%%%%%%%%%%%%%%%%%%%%%%%%%%%%%%%%%%%%%%%%%%%%%%%%%%%%%%%%%%%
%
%%%%%%%%%%%%%%%%%%%%%%%%%%%%%%%%%%%%%%%%%%%%%%%%%%%%%%%%%%%%%%%%%%%%%%%%%%%%%%%
\section{Bounds on nonlocality}
\label{sec: Bounds on nonlocality}

To begin with, let us derive the inequalities discussed in this
paper. The derivation is surprisingly simple. Consider the simplest Bell
scenario, where Alice and Bob share a quantum state, Alice (Bob) performs a
measurement depending on a given bit $x$ ($y$) to obtain the outcome bit $a$
($b$). Their shared state is a pure or mixed state denoted by $\rho$. We use
the shorthand notation $\langle \cdots \rangle$ for $\hbox{tr}\rho\cdots$.
Without loss of generality, we can assume that they perform projective
measurements, because no assumption is made about the system dimension. The
observable of Alice (Bob), denoted by $A_{x}$ ($B_{y}$), then satisfies
$A_{x}^{2}\!=\!B_{x}^{2}\!=\!I$ with $I$ being the identity operator.
The projector of the measurement for Alice's outcome $a$ is given by
$P_{a|x}\!=\!(I+(-1)^{a} A_{x})/2$, and for Bob's outcome $b$ by 
$Q_{b|y}\!=\!(I+(-1)^{b} B_{y})/2$. Let us then consider the weighted CHSH
expression \cite{Lawson10a} of the form:
%%%%%%%%%%%%%%%%%%%%%%%%%%%%%%%%%%%%%%%%%%%%%%%%%%%%%%%%%%%%%%%%%%%%%%%%%%%%%%%
\begin{equation}
{\cal B}=\sum_{y}t_y s_x (-1)^{xy} \langle A_x\!\otimes\!B_y\rangle
=\sum_y t_y E_y,
\end{equation}
%%%%%%%%%%%%%%%%%%%%%%%%%%%%%%%%%%%%%%%%%%%%%%%%%%%%%%%%%%%%%%%%%%%%%%%%%%%%%%%
where $t_y$ and $s_x$ are real non-negative parameters, and
$E_{y}\!\equiv\!s_0\langle A_0\!\otimes\!B_y\rangle+s_1(-1)^y \langle A_1\!\otimes\!B_y\rangle$ is introduced for later convenience. If we define
$X_x\!\equiv\!t_0B_0\!+\!(-1)^{x}t_1B_1$, it can be seen that
$X^{2}_0\!+\!X^{2}_1\!=\!2(t^{2}_0\!+\!t^{2}_1)I$, and we obtain the
Tsirelson-type inequality as follows:
%%%%%%%%%%%%%%%%%%%%%%%%%%%%%%%%%%%%%%%%%%%%%%%%%%%%%%%%%%%%%%%%%%%%%%%%%%%%%%%
\begin{eqnarray}
{\cal B}&\!\!\!=\!\!\!&
%\sum_{x} s_{x}\langle A_{x}\!\otimes\!X_{x}\rangle 
\sum_{x} s_{x}
\sqrt{\langle I\!\otimes\!X^{2}_{x}\rangle}
\frac{\langle A_{x}\!\otimes\!X_{x} \rangle}
{\sqrt{\langle I\!\otimes\!X^{2}_{x}\rangle}} 
\le\sum_{x}s_{x}
\sqrt{\langle I\!\otimes\!X^{2}_{x} \rangle} \tilde D_{x} \cr
&& \le \sqrt{2(t_{0}^2+t_{1}^2)(s_{0}^2\tilde D^{2}_0+s_{1}^2\tilde D^{2}_1)},
\label{eq: weighted Tsirelson bound}
\end{eqnarray}
%%%%%%%%%%%%%%%%%%%%%%%%%%%%%%%%%%%%%%%%%%%%%%%%%%%%%%%%%%%%%%%%%%%%%%%%%%%%%%%
where we used
$\sum_{x}\langle I\!\otimes\!X_{x}^{2}\rangle\!=\!2(t^{2}_0\!+\!t^{2}_1)$
as a constraint in the last inequality. The quantity $\tilde D_{x}$ is
defined by
%%%%%%%%%%%%%%%%%%%%%%%%%%%%%%%%%%%%%%%%%%%%%%%%%%%%%%%%%%%%%%%%%%%%%%%%%%%%%%%
\begin{equation}
\tilde D_{x}\equiv \max_X \frac{\langle A_{x}\!\otimes\! X \rangle}
{\sqrt{\langle I\!\otimes\!X^{2} \rangle}}
=\max_X \frac{\hbox{tr}X(\rho_{0|x}-\rho_{1|x})}{\sqrt{\hbox{tr}
X^2(\rho_{0|x}+\rho_{1|x})}},
\label{eq: tilde D}
\end{equation}
%%%%%%%%%%%%%%%%%%%%%%%%%%%%%%%%%%%%%%%%%%%%%%%%%%%%%%%%%%%%%%%%%%%%%%%%%%%%%%%
where the maximization is taken over all Hermitian operators $X$, and
$\rho_{a|x}\!=\!\hbox{tr}_A (P_{a|x}\!\otimes\!I)\rho$
is Bob's subnormalized state when Alice is given $x$ and her outcome is $a$.
The quantity $\tilde D_{x}$ is quite analogous to the generalized trace
distance $\bar D_{x}=\hbox{tr}|\rho_{0|x}-\rho_{1|x}|$ (the extension of
the trace distance to subnormalized states). Indeed, both agree with each other
for the case of pure states. For the other general cases, 
$\tilde D_{x}\!\ge\!\bar D_{x}$.
See Appendix \ref{sec: Properties of D} for the proofs of those properties.
It is obvious from the definition of
Eq.\ (\ref{eq: tilde D}) that $\tilde D_{x}\!\le\!1$, because it is
the inner product of the two normalized states
$(A_{x}\!\otimes\!I)|\psi\rangle$ and
$(I\!\otimes\!X)|\psi\rangle/\sqrt{\langle\psi| I\!\otimes\!X^2 |\psi\rangle}$
(consider a purification $|\psi\rangle$ if $\rho$ is a mixed state),
and the inner product is ensured to be real \cite{Tsirelson87a,Vertesi08a}.
Note that a different type of quantum bounds using the trace distance was
shown in \cite{Liang07a}. Since the envelope of the boundaries of
Eq.\ (\ref{eq: weighted Tsirelson bound}) in the $(E_{0},E_{1})$-space is a
quarter-circle, considering the symmetry with respect to
$B_y\!\rightarrow\!-B_y$ and putting $s_0\!=\!s_1\!=\!1/2$, we have the
IC-type inequality:
%%%%%%%%%%%%%%%%%%%%%%%%%%%%%%%%%%%%%%%%%%%%%%%%%%%%%%%%%%%%%%%%%%%%%%%%%%%%%%%
\begin{equation}
E^{2}_{0}+E^{2}_{1}
\le \frac{\tilde D^{2}_0+\tilde D^{2}_1}{2}.
\label{eq: weighted IC bound}
\end{equation}
%%%%%%%%%%%%%%%%%%%%%%%%%%%%%%%%%%%%%%%%%%%%%%%%%%%%%%%%%%%%%%%%%%%%%%%%%%%%%%%
Note that $E_{0}$ and $E_{1}$ coincides with $E_{\rm I}$ and $E_{\rm II}$
in \cite{Pawlowski09a}, respectively.

In the same technique as above, a tighter quantum bound is obtained by
considering more general weight parameters as follows:
%%%%%%%%%%%%%%%%%%%%%%%%%%%%%%%%%%%%%%%%%%%%%%%%%%%%%%%%%%%%%%%%%%%%%%%%%%%%%%%
\begin{equation}
\sum_{xy}s_x u_{xy}(-1)^{xy}\langle A_x\otimes B_y\rangle
\le\big[\sum_{xy}u^{2}_{xy}\big]^{\frac{1}{2}}
\big[\sum_{x}s^{2}_x \tilde D^{2}_x \big]^{\frac{1}{2}},
\label{eq: weighted CHSH bound}
\end{equation}
%%%%%%%%%%%%%%%%%%%%%%%%%%%%%%%%%%%%%%%%%%%%%%%%%%%%%%%%%%%%%%%%%%%%%%%%%%%%%%%
where $s_{x}$ and $u_{xy}$ are real parameters satisfying
$u_{00}u_{01}\!=\!u_{10}u_{11}$. When $\tilde D_0,\tilde D_1\!>\!0$, the
necessary and sufficient condition for the above inequality is given by
%%%%%%%%%%%%%%%%%%%%%%%%%%%%%%%%%%%%%%%%%%%%%%%%%%%%%%%%%%%%%%%%%%%%%%%%%%%%%%%
\begin{eqnarray}
\left|\tilde C_{00}\tilde C_{01}-\tilde C_{10}\tilde C_{11}\right|
&\le& (1-\tilde C^{2}_{00})^{\frac{1}{2}}
(1-\tilde C^{2}_{01})^{\frac{1}{2}} \cr
&+&(1-\tilde C^{2}_{10})^{\frac{1}{2}}(1-\tilde C^{2}_{11})^{\frac{1}{2}},
\label{eq: extended Landau bound}
\end{eqnarray}
%%%%%%%%%%%%%%%%%%%%%%%%%%%%%%%%%%%%%%%%%%%%%%%%%%%%%%%%%%%%%%%%%%%%%%%%%%%%%%%
where $\tilde C_{xy}\!\equiv\!\langle A_x\otimes B_y\rangle/\tilde D_x
\!\equiv\!C_{xy}/\tilde D_x$.
The derivation is given in Appendix \ref{sec: Landau-type inequality}.
This is an extension of the Landau inequality \cite{Landau88a}.
The Landau inequality is a representation of the TLM criterion, and hence is
necessary and sufficient so that a given set of the conditional probabilities
$\{p(ab|xy)\}$ (or a given set of $\{C_{xy}\}$) is quantum realizable in the
case of unbiased marginals such that $p(a|x)\!=\!p(b|y)\!=\!1/2$.
It is known that the Navascu\'es-Pironio-Ac\'{\i}n (NPA) inequality
\cite{Navascues07a,Navascues08a} gives a tighter bound than the Landau
inequality for the case of biased marginals. It is also possible to extend the
NPA inequality to include $\tilde D_x$ as shown in
Appendix \ref{sec: Landau-type inequality}.

The above inequalities all represent the effects of the nonorthogonality
between Bob's local states for $a\!=\!0$ and 1. Indeed, when $\rho_{0|x}$ and
$\rho_{1|x}$ are orthogonal for both $x=0$ and $1$, we have
$\tilde D_0\!=\!\tilde D_1\!=\!1$, and those reproduce the inequalities known
so far. Note that $\tilde D_x\!=\!1$ if and only if $\bar D_x\!=\!1$ (see
Appendix \ref{sec: Properties of D}) and so also in the case of
$\hbox{tr}\rho_{0|x}\!=\!0$ or $\hbox{tr}\rho_{1|x}\!=\!0$.
This is included in the orthogonal case throughout this paper.

%%%%%%%%%%%%%%%%%%%%%%%%%%%%%%%%%%%%%%%%%%%%%%%%%%%%%%%%%%%%%%%%%%%%%%%%%%%%%%%
%
%%%%%%%%%%%%%%%%%%%%%%%%%%%%%%%%%%%%%%%%%%%%%%%%%%%%%%%%%%%%%%%%%%%%%%%%%%%%%%%
\section{Tightness of bounds}
\label{sec: Tightness of bounds}

The inequalities derived in Sec.\ \ref{sec: Bounds on nonlocality} must hold
for all physical realizations (by projective measurements). A nonlocal
correlation is generally identified by the set of conditional probabilities
$\{p(ab|xy)\}$, and the left-hand side of e.g.,
Eq.\ (\ref{eq: weighted CHSH bound})
is determined by $\{p(ab|xy)\}$ only (for a fixed weight). On the other hand,
the right-hand side is monotonically increasing with respect to $\tilde D_x$.
It is then found that, if a set $\{p(ab|xy)\}$ saturates the Landau
inequality [i.e., the equality of Eq.\ (\ref{eq: weighted CHSH bound}) holds
with $\tilde D_0\!=\!\tilde D_1\!=\!1$ by appropriately chosen weight
parameters], the realization that produces the same $\{p(ab|xy)\}$
but with $\tilde D_x\!<\!1$ is not allowed. Namely, we have the following:

{\bf Lemma 1.}
{\it For every correlation that saturates the Landau inequality, there
is no realization such that Bob's subnormalized states $\rho_{0|x}$ and
$\rho_{1|x}$ are nonorthogonal. The same holds for Alice's states.
}

Note that this is the case of the NPA inequality by
Eq.\ (\ref{eq: weighted tilted CHSH bound}). Note further that Lemma 1 is
consistent with the fact that the nonclassical boundary correlations with
unbiased marginals are all used for the self-testing of the maximally entangled
state of two qubits, i.e., solely realized by the maximally entangled
state \cite{Wang16a}, because every boundary correlation with unbiased
marginals is given by the saturation of the Landau inequality.

In the case of the Landau-type inequality
Eq.\ (\ref{eq: extended Landau bound}) that includes $\tilde D_x$, the
saturation does not necessarily imply that the correlation is located at a
boundary. Rather, the inequality is widely saturated even for non-boundary
correlations. To see this, let us consider the completely random correlation
$\mathbf{I}$ given by $p(ab|xy)=1/4$ ($\forall a,b,x,y$), which is realized by
the maximally mixed state of two qubits
$\rho_{AB}\!=\!\frac{1}{4}I_A\!\otimes\!I_B$, where $C_{xy}\!=\!0$ and
$\tilde D_x\!=\!0$. Then, if Eq.\ (\ref{eq: extended Landau bound}) is
saturated for a correlation $\mathbf{p}$ by some realization, it is also done
for $\mathbf{q}$ of the form
%%%%%%%%%%%%%%%%%%%%%%%%%%%%%%%%%%%%%%%%%%%%%%%%%%%%%%%%%%%%%%%%%%%%%%%%%%%%%%%
\begin{equation}
\mathbf{q}\!=\!\lambda \mathbf{p}+(1-\lambda)\mathbf{I},
\label{eq: convex combination}
\end{equation}
%%%%%%%%%%%%%%%%%%%%%%%%%%%%%%%%%%%%%%%%%%%%%%%%%%%%%%%%%%%%%%%%%%%%%%%%%%%%%%%
where $0\!\le\!\lambda\!\le\!1$.
This is because, when $\mathbf{p}$ is realized by $\rho^{\mathbf{p}}_{AB}$,
$\mathbf{q}$ is realized by the shared state of
%%%%%%%%%%%%%%%%%%%%%%%%%%%%%%%%%%%%%%%%%%%%%%%%%%%%%%%%%%%%%%%%%%%%%%%%%%%%%%%
\begin{equation}
\rho^{\mathbf{p}}_{AB}\otimes({\textstyle \frac{1}{4}}I_{A}\otimes I_{B})
\otimes
\big[\lambda|00\rangle\langle00|+(1-\lambda)|11\rangle\langle11|\big]_{AB}
\label{eq: shared state}
\end{equation}
%%%%%%%%%%%%%%%%%%%%%%%%%%%%%%%%%%%%%%%%%%%%%%%%%%%%%%%%%%%%%%%%%%%%%%%%%%%%%%%
such that Alice and Bob switch their measured states (and the corresponding
measurements) between
$\rho^{\mathbf{p}}_{AB}$ and $\frac{1}{4}(I_{A}\!\otimes\!I_{B})$ according to
the shared randomness produced by
$\lambda|00\rangle\langle00|\!+\!(1\!-\!\lambda)|11\rangle\langle11|$,
and it is found from the closed form of $\tilde D_x$ (see 
Appendix \ref{sec: Properties of D}) that $\tilde D_{x}$
for $\mathbf{q}$ and $\mathbf{p}$
are related through
$\tilde D^{\mathbf{q}}_{x}\!=\!\lambda \tilde D^{\mathbf{p}}_{x}
\!+\!(1-\lambda) \tilde D^{\mathbf{I}}_{x}
\!=\!\lambda \tilde D^{\mathbf{p}}_{x}$, and hence
$\tilde C^{\mathbf{q}}_{xy}\!=\!\tilde C^{\mathbf{p}}_{xy}$ holds.
This implies that, if Eq.\ (\ref{eq: extended Landau bound}) is saturated for
every boundary of the set of quantum correlations, the inequality is saturated
for all correlations inside the set. This is indeed the case of unbiased
marginals, because the inequality is saturated for every boundary
with $\tilde D_0\!=\!\tilde D_1\!=\!1$, and we obtain the following:

{\bf Lemma 2.}
{\it For every correlation with unbiased marginals, there always exists a
realization such that the equality holds in
Eq.\ (\ref{eq: extended Landau bound}).
}

An important observation is that the inequality is saturated
even for the case of biased marginals. A two-qubit realization to give
the maximal violation of the Bell expression
$\beta\langle A_0\rangle+\alpha\langle A_0B_0\rangle
+\alpha\langle A_0B_1\rangle
+\langle A_1B_0\rangle
-\langle A_1B_1\rangle$ was shown in \cite{Acin12a}, where the partially 
entangled state $|\psi\rangle=\cos\theta|00\rangle+\sin\theta|11\rangle$
produces the boundary correlations with biased marginals.
In this realization, we have
$\tilde D_0\!=\!1$ and $\tilde D_1\!=\!\sin2\theta$ irrespective of $\alpha$,
$\tilde C_{00}\!=\!\tilde C_{01}\!=\!
\frac{\alpha}{\sqrt{\sin^2 2\theta+\alpha^2}}$,
$\tilde C_{10}\!=\!-\tilde C_{11}\!=\!
\frac{\sin2\theta}{\sqrt{\sin^2 2\theta+\alpha^2}}$, and
Eq.\ (\ref{eq: extended Landau bound}) is saturated for a whole range of
$\alpha$ and $\sin2\theta$.

It is known that any extremal nonclassical correlation in the simplest
Bell scenario has a two-qubit realization, where projective measurements of
rank 1 are performed on a pure entangled state \cite{Masanes06a}. For such
extremal realizations, by applying appropriate local unitary
transformations, Alice and Bob's observables are written as
%%%%%%%%%%%%%%%%%%%%%%%%%%%%%%%%%%%%%%%%%%%%%%%%%%%%%%%%%%%%%%%%%%%%%%%%%%%%%%%
\begin{equation}
A_x=\cos \theta^{A}_x \sigma_1 + \sin \theta^{A}_x \sigma_3,\hbox{~}
B_y=\cos \theta^{B}_y \sigma_1 + \sin \theta^{B}_y \sigma_3,
\label{eq: Parametrization 1}
\end{equation}
%%%%%%%%%%%%%%%%%%%%%%%%%%%%%%%%%%%%%%%%%%%%%%%%%%%%%%%%%%%%%%%%%%%%%%%%%%%%%%%
where $(\sigma_1,\sigma_3,\sigma_3)$ are the Pauli matrices. Moreover,
$\rho\!=\!|\psi\rangle\langle\psi|$ is chosen to be real symmetric, and let
us express
%%%%%%%%%%%%%%%%%%%%%%%%%%%%%%%%%%%%%%%%%%%%%%%%%%%%%%%%%%%%%%%%%%%%%%%%%%%%%%%
\begin{eqnarray}
\!\!\!\!\!\!\!\!\hbox{tr}_A (A_x\!\otimes\!I) |\psi\rangle\langle\psi|
&\!\!=\!\!&\alpha^{B}_x I +
\beta^{B}_x(\cos \phi^{B}_x \sigma_1 + \sin \phi^{B}_x \sigma_3), \cr
\!\!\!\!\!\!\!\!\hbox{tr}_B (I\!\otimes\!B_y) |\psi\rangle\langle\psi|
&\!\!=\!\!&\alpha^{A}_y I + 
\beta^{A}_y(\cos \phi^{A}_y \sigma_1 + \sin \phi^{A}_y \sigma_3).
\label{eq: Parametrization 2}
\end{eqnarray}
%%%%%%%%%%%%%%%%%%%%%%%%%%%%%%%%%%%%%%%%%%%%%%%%%%%%%%%%%%%%%%%%%%%%%%%%%%%%%%%
As shown in Appendix \ref{sec: Extremal correlation}, the necessary and
sufficient condition for the saturation of
Eq.\ (\ref{eq: extended Landau bound}) is given by
%%%%%%%%%%%%%%%%%%%%%%%%%%%%%%%%%%%%%%%%%%%%%%%%%%%%%%%%%%%%%%%%%%%%%%%%%%%%%%%
\begin{equation}
\sin(\phi^{B}_0-\theta^{B}_0)\sin(\phi^{B}_0-\theta^{B}_1)
\sin(\phi^{B}_1-\theta^{B}_0)\sin(\phi^{B}_1-\theta^{B}_1)\le0.
\label{eq: Saturation condition B}
\end{equation}
%%%%%%%%%%%%%%%%%%%%%%%%%%%%%%%%%%%%%%%%%%%%%%%%%%%%%%%%%%%%%%%%%%%%%%%%%%%%%%%
Similarly, for the counterpart inequality on
Alice's side,
%%%%%%%%%%%%%%%%%%%%%%%%%%%%%%%%%%%%%%%%%%%%%%%%%%%%%%%%%%%%%%%%%%%%%%%%%%%%%%%
\begin{equation}
\sin(\phi^{A}_0-\theta^{A}_0)\sin(\phi^{A}_0-\theta^{A}_1)
\sin(\phi^{A}_1-\theta^{A}_0)\sin(\phi^{A}_1-\theta^{A}_1)\le0.
\label{eq: Saturation condition A}
\end{equation}
%%%%%%%%%%%%%%%%%%%%%%%%%%%%%%%%%%%%%%%%%%%%%%%%%%%%%%%%%%%%%%%%%%%%%%%%%%%%%%%
We have performed the Monte Carlo calculations, where a two-qubit realization
to give the maximal violation of a randomly generated Bell expression is
obtained. The numerical results suggest that
Eq.\ (\ref{eq: Saturation condition B})
and (\ref{eq: Saturation condition A}) are simultaneously satisfied for all
nonclassical extremal correlations, and hence support the following
conjecture:

{\bf Conjecture 1.}
{\it For every extremal correlation, there always exists a realization such
that the equality holds in Eq.\ (\ref{eq: extended Landau bound}) and in
the counterpart inequality on Alice's side.
}

In this way, most of correlations including the case of biased marginals
appear to obey a simple and unified rule, which is revealed by
considering the nonorthogonality between local states. Note that, when the
real symmetric
$\rho$ is maximally entangled, Eq.\ (\ref{eq: Saturation condition B}) and
(\ref{eq: Saturation condition A}), which specify a geometric relation
between angles (see \cite{Wang16a}), are necessary and sufficient for the
extremality of the generated correlation with unbiased marginals.
When the real symmetric $\rho$ is chosen to be pure and partially entangled, do
those provide a simple necessary and sufficient condition for the extremality
also in the case of biased marginals? This is an intriguing open problem.

%%%%%%%%%%%%%%%%%%%%%%%%%%%%%%%%%%%%%%%%%%%%%%%%%%%%%%%%%%%%%%%%%%%%%%%%%%%%%%%
%
%%%%%%%%%%%%%%%%%%%%%%%%%%%%%%%%%%%%%%%%%%%%%%%%%%%%%%%%%%%%%%%%%%%%%%%%%%%%%%%
\section{Information theoretical aspects}
\label{sec: Information theoretical aspects}

\begin{figure}[t]
\centerline{\scalebox{0.45}[0.45]{\includegraphics{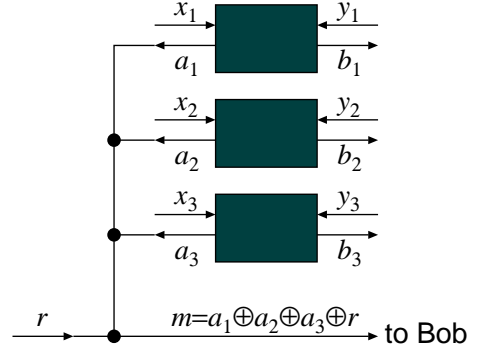}}}
\caption{
The case of $n\!=\!3$ is shown. The parity bit of all the outputs of the $n$
boxes is scrambled by a random bit $r$ and is sent to Bob as a message.
}
\label{fig: PWIC protocol}
\end{figure}

To discuss the information theoretical aspects of the IC-type inequality
Eq.\ (\ref{eq: weighted IC bound}), let us introduce a communication protocol.
A nonlocal game known as the inner product game has been studied in connection
to communication complexity \cite{Buhrman10a,vanDam13a}. The protocol we
consider is its communication version shown in Fig.\ \ref{fig: PWIC protocol},
where Alice and Bob is given a random $n$-bit string
$\vec x\!=\!(x_1,\cdots,x_n)$ and $\vec y\!=\!(y_1,\cdots,y_n)$
generated with the probability $s_{\vec x}$ and $t_{\vec y}$, respectively.
Alice (Bob) outputs a bit $a_p$ ($b_p$) utilizing shared quantum states,
and she sends the message $m$ to Bob that is $a_p$ scrambled by an independent
random bit $r$ such as $m=a_p\oplus r$. The purpose of this protocol is that
Bob obtains the value of $(\vec x\cdot\vec y)\oplus r$, where
$\vec x\cdot\vec y\!=\!\sum_i x_i y_i \mod 2$. The task is nontrivial
even for $\vec y=(0,\cdots,0)$ due to the scrambling by $r$.
A more important role of $r$ becomes clear later.
Let $A_{\vec x}$ ($B_{\vec y}$) be the observable of Alice (Bob) to obtain
$a_p$ ($b_p$), and the projector of Alice (Bob) be $P_{a_p|\vec x}$
($Q_{b_p|\vec y}$). Then, Bob's success probability for a given $\vec y$
averaged over $\vec x$ is 
%%%%%%%%%%%%%%%%%%%%%%%%%%%%%%%%%%%%%%%%%%%%%%%%%%%%%%%%%%%%%%%%%%%%%%%%%%%%%%%
\begin{eqnarray}
p_{\vec y}&=&\sum_{\vec x}s_{\vec x}\sum_{a_p b_p}
\langle P_{a_p|\vec x}\otimes Q_{b_p|\vec y} \rangle
\delta_{\vec x\cdot\vec y=a_p\oplus b_p} \cr
%&=&\sum_{\vec x}s_{\vec x}
%\langle
%{\textstyle \frac{I\otimes I+A_{\vec x}\otimes B_{\vec y}}{2}}
%\delta_{\vec x\cdot\vec y=0}
%+{\textstyle \frac{I\otimes I-A_{\vec x}\otimes B_{\vec y}}{2}}
%\delta_{\vec x\cdot\vec y=1}\rangle \cr
&=&\frac{1}{2}\big(1+
\sum_{\vec x} s_{\vec x}\langle A_{\vec x}\otimes (-1)^{\vec x \cdot\vec y}B_{\vec y}\rangle\big).
\end{eqnarray}
%%%%%%%%%%%%%%%%%%%%%%%%%%%%%%%%%%%%%%%%%%%%%%%%%%%%%%%%%%%%%%%%%%%%%%%%%%%%%%%
Concerning the quantum bound of the bias
$E_{\vec y}\!\equiv\!2p_{\vec y}\!-\!1$, the discussion runs in parallel with
Sec.\ \ref{sec: Bounds on nonlocality}. Indeed, 
$\sum_{\vec x}X_{\vec x}^{2}\!=\!2^{n}\sum_{\vec y}t_{\vec y}^2 I$ holds for
$X_{\vec x}\!\equiv\!\sum_{\vec y}t_{\vec y}(-1)^{\vec x\cdot\vec y}B_{\vec y}$,
hence
$\sum_{\vec y}t_{\vec y}E_{\vec y}\!\le\!(2^n\sum_{\vec x\vec y}
t_{\vec y}^2s_{\vec x}^2 \tilde D^{2}_{\vec x})^{\frac{1}{2}}$, and we have
$\sum_{\vec y}E^{2}_{\vec y}\!\le\!2^n\sum_{\vec x}s_{\vec x}^2\tilde D^{2}_{\vec x}$.
Let us now assume that Alice and Bob utilize the $n$
identical ``quantum boxes'', each of which accepts inputs $(x,y)$
and produces outputs $(a,b)$ according to $\{p(ab|xy)\}$,
and assume that $a_p$ ($b_p$) is the parity bit of Alice's
(Bob's) outputs from the $n$ boxes as shown in Fig.\ \ref{fig: PWIC protocol}.
Under those assumptions, $B_{\vec y}$ must have a tensor product form such as
$B_{\vec y}\!=\!B_{y_1}\otimes B_{y_2}\otimes\cdots$,
which implies that the maximization operator $X$ in $\tilde D_{\vec x}$ also
has a tensor product form. It is then found that
%%%%%%%%%%%%%%%%%%%%%%%%%%%%%%%%%%%%%%%%%%%%%%%%%%%%%%%%%%%%%%%%%%%%%%%%%%%%%%%
\begin{equation}
\sum_{\vec y}E^{2}_{\vec y}\le\big(\frac{\tilde D^{2}_0+\tilde D^{2}_1}{2}\big)^n
\label{eq: large n bound}
\end{equation}
%%%%%%%%%%%%%%%%%%%%%%%%%%%%%%%%%%%%%%%%%%%%%%%%%%%%%%%%%%%%%%%%%%%%%%%%%%%%%%%
must hold in this protocol for $s_{\vec x}\!=\!1/2^n$, whose right-hand side is
the $n$-th power of the right-hand side of Eq.\ (\ref{eq: weighted IC bound}).

In the general setting of communication, where Alice is given $\vec x$ and
sends the bit string $\vec m$ to Bob as a message, the information obtainable
by Bob is characterized by the mutual information $I(\vec x\!:\!\vec m\rho_B)$,
where $\rho_B$ is the state of Bob's half of no-signaling resources. Using the
no-signaling condition and the information-theoretical relations respected by
quantum mechanics, it was shown that \cite{Pawlowski09a,Dahlsten13a,Al-Safi11a}
%%%%%%%%%%%%%%%%%%%%%%%%%%%%%%%%%%%%%%%%%%%%%%%%%%%%%%%%%%%%%%%%%%%%%%%%%%%%%%%
\begin{eqnarray}
I(\vec x:\vec m \rho_B) 
&=&I(\vec m:\vec x \rho_B)-I(\vec m:\rho_B) \cr
&\le& H(\vec m)-H(\vec m|\vec x \rho_B) \le H(\vec m).
\label{eq: IC principle}
\end{eqnarray}
%%%%%%%%%%%%%%%%%%%%%%%%%%%%%%%%%%%%%%%%%%%%%%%%%%%%%%%%%%%%%%%%%%%%%%%%%%%%%%%
Since the entropy $H(\vec m)$ cannot exceed the number of bits in $\vec m$,
the IC principle is derived. The left-hand side of
Eq.\ (\ref{eq: large n bound}), where the $2^n$ variables Bob tries to obtain
are pair-wise independent \cite{Pawlowski12a,Al-Safi11a,Bavarian15a},
generally corresponds to the term $I(\vec x\!:\!\vec m \rho_B)$. To investigate
the origin of the right-hand side of Eq.\ (\ref{eq: large n bound}),
let us focus on the term $H(\vec m|\vec x\rho_B)$ omitted in the
derivation of the IC principle (also in a generalization of the IC
inequality \cite{Chaves14a}).

%%%%%%%%%%%%%%%%%%%%%%%%%%%%%%%%%%%%%%%%%%%%%%%%%%%%%%%%%%%%%%%%%%%%%%%%%%%%%%
%%%%%%%%%%%%%%%%%%%%%%%%%%%%%%%%%%%%%%%%%%%%%%%%%%%%%%%%%%%%%%%%%%%%%%%%%%%%%%%

In the protocol of Fig.\ \ref{fig: PWIC protocol}, since Alice is given
$\vec x$ and $r$, the relation corresponding to
Eq.\ (\ref{eq: IC principle}) is
%%%%%%%%%%%%%%%%%%%%%%%%%%%%%%%%%%%%%%%%%%%%%%%%%%%%%%%%%%%%%%%%%%%%%%%%%%%%%%%
\begin{equation}
I(\vec x r:m\rho_B) = H(m)-H(m|\vec x r\rho_B)=1-H(a_p|\vec x\rho_B),
\label{eq: PWIC principle}
\end{equation}
%%%%%%%%%%%%%%%%%%%%%%%%%%%%%%%%%%%%%%%%%%%%%%%%%%%%%%%%%%%%%%%%%%%%%%%%%%%%%%%
where we took into account $I(m\!:\!\rho_B)\!=\!0$ and used
$H(m|\vec x r\rho_B)\!=\!H(a_p|\vec x\rho_B)$ because the conditional entropy
$H(X|Y)\!\equiv\!H(XY)\!-\!H(Y)$ means the remaining uncertainty of $X$ after
knowing $Y$ (in the classical variable case). Let us then evaluate
$1-H(a_p|\vec x\rho_B)$ in quantum mechanics. Considering the individual
measurement strategy for boxes, the optimal success probability of guessing
Alice's outcome $a$ of a single box for the input $x$ is given by
$(1\!+\!\bar D_x)/2$, which is an operational meaning of the generalized trace
distance. As shown in Appendix \ref{sec: Evaluation of 1-H},
the result of the evaluation in the $n\!\rightarrow\!\infty$ limit is then
%%%%%%%%%%%%%%%%%%%%%%%%%%%%%%%%%%%%%%%%%%%%%%%%%%%%%%%%%%%%%%%%%%%%%%%%%%%%%%%
\begin{equation}
1-H(a_p|\vec x\rho_B) = \frac{1}{2\ln2} \big(\frac{\bar D^{2}_0+\bar D^{2}_1}{2}\big)^n,
\label{eq: Bounds of 1-H}
\end{equation}
%%%%%%%%%%%%%%%%%%%%%%%%%%%%%%%%%%%%%%%%%%%%%%%%%%%%%%%%%%%%%%%%%%%%%%%%%%%%%%%
which appears to well correspond to the right-hand side of
Eq.\ (\ref{eq: large n bound}) (although there is a slight difference between
$\tilde D_x$ and $\bar D_x$ in the case of mixed states).

In this way, it is found that the inequalities discussed in this paper
represent the effects of the nonzero $H(\vec m|\vec x\rho_B)$ in
Eq.\ (\ref{eq: IC principle}). For the nonzeroness, it is crucial whether
or not Bob can completely determine Alice's outcome (abstractly denoted by
$\vec a$ hereafter) from the type of her measurement $\vec x$ and his local
state $\rho_B$. If he cannot do this, it
implies $H(\vec a|\vec x\rho_B)\!>\!0$ and results in
$H(\vec m|\vec x\rho_B)\!>\!0$ when $\vec m$ is constructed from $\vec a$
and $\vec x$. In this situation, $\vec a$ appears to
have some randomness and be scrambling the information of $\vec x$ encoded in
$\vec m$ from the viewpoint of Bob.
This can occur not only when quantum resources are mixed states, but also
pure states. Indeed, quantum correlations, which can be realized
by partially entangled states (whose Schmidt coefficients are nondegenerate 
so that the Schmidt basis is unique), inevitably show
$H(\vec a|\vec x\rho_B)\!>\!0$,
because Alice's measurements are non-commuting \cite{Fine82a} and the basis of
at least one measurement differs from the Schmidt basis. As a result, Bob's
local states for different values of $\vec a$ become nonorthogonal, and he
cannot completely determine $\vec a$. It is a peculiar feature of quantum
mechanics that there exist the extremal correlations that are realized by
partially entangled states \cite{Liang11a,Acin12a,Ramanathan15a,Zhen16a} and
show $H(\vec a|\vec x\rho_B)\!>\!0$, because every extremal correlation
of both sets of classical and general no-signaling correlations (local
deterministic correlations \cite{Brunner14a} and the Popescu-Rohrlich type
boxes \cite{Popescu94a,Barrett05a}) does not show
$H(\vec a|\vec x\rho_B)\!>\!0$.

The randomness discussed above inevitably reduces the information obtainable by
Bob. Indeed, it is clear from Eq.\ (\ref{eq: IC principle}) that, for a quantum
correlation that shows nonzero $H(\vec a|\vec x\rho_B)$, any protocol whose
$\vec m$ contains the information of $\vec a$ and
$H(\vec m|\vec x\rho_B)\!>\!0$ cannot achieve
$I(\vec x\!:\!\vec m\rho_B)\!=\!H(\vec m)$ (the achievement is possible when
$\vec m$ does not contain the information of $\vec a$, but in that case the
quantum correlation is not used by the protocol). This include the
case of the extremal correlations realized by partially entangled states
discussed above. For those nonlocal correlations, 
the strength is constrained by a principle other than the IC principle. To
investigate what the principle is, let us consider the protocol of
Fig.\ \ref{fig: PWIC protocol} again. The point is that the message $m$ is
completely scrambled by $r$. As a result, $I(\vec x:m\rho_B)=0$ must hold
by the cryptographic principle (or the principle of the information theoretic
security), which states that a completely scrambled message (i.e., scrambled by
independent random bits with the same number of the message bits) cannot carry
information \cite{Shannon49a}. This cryptographic principle is derived in the
same way as in \cite{Pawlowski09a} using the chain rule of mutual information
[$I(A\!:\!B|C)=I(A\!:\!BC)-I(A\!:\!C)$] and the exchange symmetry
[$I(A\!:\!B|C)=I(B\!:\!A|C)$] as
%%%%%%%%%%%%%%%%%%%%%%%%%%%%%%%%%%%%%%%%%%%%%%%%%%%%%%%%%%%%%%%%%%%%%%%%%%%%%%%
\begin{eqnarray}
I(\vec x:m\rho_B)&=&I(\vec x:\rho_B|m)+I(\vec x:m) \cr
&=&I(\rho_B:\vec x m)-I(\rho_B:m)=0,
\end{eqnarray}
%%%%%%%%%%%%%%%%%%%%%%%%%%%%%%%%%%%%%%%%%%%%%%%%%%%%%%%%%%%%%%%%%%%%%%%%%%%%%%%
where we used $I(\vec x m:\rho_B)=0$ by the no-signaling condition and used
$I(m\!:\!\vec x)\!=\!I(m\!:\!\rho_B)\!=\!0$ by the independence
of $r$. From this cryptographic principle, Eq.\ (\ref{eq: PWIC principle}) is
also obtained as
%%%%%%%%%%%%%%%%%%%%%%%%%%%%%%%%%%%%%%%%%%%%%%%%%%%%%%%%%%%%%%%%%%%%%%%%%%%%%%%
\begin{eqnarray}
0&=&I(\vec x:m\rho_B)=I(m\rho_B:\vec x r)-I(m\rho_B:r|\vec x) \cr
&=& I(\vec x r:m\rho_B)-I(r:\vec x m\rho_B) \cr
&=& I(\vec x r:m\rho_B)-I(m:\vec x r\rho_B) \cr
&=& I(\vec x r:m\rho_B)-1+H(a_p|\vec x\rho_B),
\label{eq: Cryptographic causality}
\end{eqnarray}
%%%%%%%%%%%%%%%%%%%%%%%%%%%%%%%%%%%%%%%%%%%%%%%%%%%%%%%%%%%%%%%%%%%%%%%%%%%%%%%
where the independence of $r$ was again used. From this, it is found that 
$I(\vec x r\!:\!m\rho_B)\!>\!1-H(a_p|\vec x\rho_B)$ implies
$I(\vec x\!:\!m\rho_B)\!>\!0$;
the transmission of
information of $\vec x r$ beyond $1-H(a_p|\vec x\rho_B)$ implies that the
completely scrambled message $m$ would carry information of $\vec x$.
Namely, the strength of the nonlocal correlation is constrained so that the
violation of the cryptographic principle does not occur.
The inequalities derived in this paper represent the effects of nonzero
$H(a_p|\vec x\rho_B)$, and those can be said to be cryptographic quantum
bounds on nonlocality.

%%%%%%%%%%%%%%%%%%%%%%%%%%%%%%%%%%%%%%%%%%%%%%%%%%%%%%%%%%%%%%%%%%%%%%%%%%%%%%%
%
%%%%%%%%%%%%%%%%%%%%%%%%%%%%%%%%%%%%%%%%%%%%%%%%%%%%%%%%%%%%%%%%%%%%%%%%%%%%%%%
\section{Boundary condition}
\label{sec: Boundary condition}

%%%%%%%%%%%%%%%%%%%%%%%%%%%%%%%%%%%%%%%%%%%%%%%%%%%%%%%%%%%%%%%%%%%%%%%%%%%%%%
\begin{figure}[t]
\centerline{\scalebox{0.35}[0.35]{\includegraphics{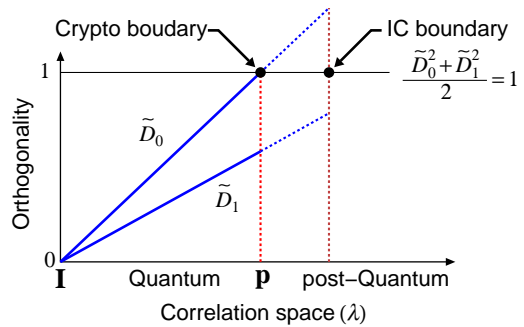}}}
\caption{
The variation of $\tilde D_0$ and $\tilde D_1$ as a function of $\lambda$
in the correlation space of
$\lambda \mathbf{p}+(1-\lambda)\mathbf{I}$, where $\mathbf{p}$ is a boundary correlation.
}
\label{fig: C boundary}
\end{figure}
%%%%%%%%%%%%%%%%%%%%%%%%%%%%%%%%%%%%%%%%%%%%%%%%%%%%%%%%%%%%%%%%%%%%%%%%%%%%%%%

The information theoretical relation representing the cryptographic principle
is an equality such as Eq.\ (\ref{eq: Cryptographic causality}), which is
consistent with the fact that the equality of
Eq.\ (\ref{eq: extended Landau bound}) widely holds not only for boundary
correlations but also for non-boundary correlations. However, this is an
undesirable property for the purpose of identifying the quantum boundaries.
Nevertheless, the cryptographic bounds tell us a way to determine the
boundaries. The two-qubit realization shown in \cite{Acin12a} and discussed in
Sec.\ \ref{sec: Tightness of bounds} again gives an informative
example. Consider the boundary correlation that maximally violates the Bell
expression
$\frac{2}{\sqrt{3}}\langle A_0\rangle+\langle A_0B_0\rangle
+\langle A_0B_1\rangle
+\langle A_1B_0\rangle
-\langle A_1B_1\rangle$, where
$|\psi\rangle\!=\!\sqrt{2/3}|00\rangle\!+\!\sqrt{1/3}|11\rangle$,
$\langle A_0\rangle\!=\!1/3$,
$\langle A_1\rangle\!=\!0$,
$\langle B_0\rangle\!=\!\langle B_1\rangle\!=\!1/\sqrt{17}$,
$\langle A_0B_0\rangle\!=\!\langle A_0B_1\rangle\!=\!3/\sqrt{17}$,
$\langle A_1B_0\rangle\!=\!-\langle A_1B_1\rangle\!=\!8/\sqrt{17}/3$,
$\tilde D_0\!=\!1$, and $\tilde D_1\!=\!\sqrt{8}/3$. This is the same as the
simple example shown in Sec.\ \ref{sec: Introduction}. For this boundary
correlation, the IC inequality is not saturated as
$E^{2}_{0}\!+\!E^{2}_{1}\!=\!17/18\!<\!1$, while the cryptographic bound of the
IC-type Eq.\ (\ref{eq: weighted IC bound}) is saturated as
$(\tilde D^{2}_0\!+\!\tilde D^{2}_1)/2\!=\!17/18$.
Note that, since the left-hand side is the same for both inequalities, the
saturation of the IC-type inequality implies that the protocol used for the
derivation of the IC inequality in \cite{Pawlowski09a} is already optimal for
maximizing the left-hand side.

Let us then
consider the correlation of the form Eq.\ (\ref{eq: convex combination}) with
${\mathbf p}$ being the boundary correlation. As discussed in
Sec.\ \ref{sec: Tightness of bounds}, $\tilde D_0$ and $\tilde D_1$ of
${\mathbf q}$ that saturate Eq.\ (\ref{eq: extended Landau bound}) vary
linearly with $\lambda$ as schematically shown in Fig.\ \ref{fig: C boundary}. 
It is then found that the quantum boundary is determined such that
$\tilde D_0$ reaches the maximum limit of 1. Indeed, if $\tilde D_0$ only takes
1 over all possible realizations of a correlation, the correlation must be
located at a boundary, because, if ${\mathbf q}$
with $\lambda\!=\!\lambda_0\!>\!1$ is quantum realizable, ${\mathbf p}$ has a
realization with $\tilde D_{0}\!=\!1/\lambda_0\!<\!1$, which causes a
contradiction. This is indeed the case of ${\mathbf p}$ because the
realization was shown to be unique up to local unitary
transformations \cite{Acin12a}.

In this way, $\tilde D_0$ and $\tilde D_1$, which must not exceed 1,
individually set a limit to determine the quantum boundaries. Every boundary
in the case of unbiased marginals can be identified in
such a way by Lemma 1. This is the case of local deterministic correlations
also, where either $\hbox{tr}\rho_{0|x}\!=\!0$ or $\hbox{tr}\rho_{1|x}\!=\!0$
holds and there is no realization with $\tilde D_x\!<\!1$. Unfortunately,
however, the results of the Monte Carlo calculations indicate that
both $\tilde D_0$ and $\tilde D_1$ (and Alice's counterparts) are generally
less than 1 for extremal correlations with biased marginals, i.e.,
most are determined by another limit, in spite that the equality of
Eq.\ (\ref{eq: extended Landau bound}) is respected. What is the principle to
fully identify the boundaries? This still remains open.

%%%%%%%%%%%%%%%%%%%%%%%%%%%%%%%%%%%%%%%%%%%%%%%%%%%%%%%%%%%%%%%%%%%%%%%%%%%%%%%
%
%%%%%%%%%%%%%%%%%%%%%%%%%%%%%%%%%%%%%%%%%%%%%%%%%%%%%%%%%%%%%%%%%%%%%%%%%%%%%%%
\section{Summary}
\label{sec: Summary} 

To conclude, we obtained the nonlocality inequalities in the simplest
Bell scenario, which must be respected by quantum mechanics and include the
effects of the randomness produced in the message when quantum resources such
as partially entangled states and mixed states are used for communication.
The randomness originates from the nonorthogonality of receiver's states and
the effects enter the inequalities through the trace distancelike quantity,
which is hence close to the bias of the optimal success probability of guessing
the sender's measurement outcome, when assuming that a receiver knows the type
of the measurement. The obtained inequalities reflect the constraint by the
cryptographic principle. This is due to the fact that the randomness reduces
the information obtainable by a receiver, and the transmission of information
beyond the reduction implies that a completely scrambled message would carry
information.

Introducing the cryptographic principle to nonlocality inequalities leads to
two effects. First, the inequalities come to be saturated
inside the set of quantum correlations. Indeed, the obtained Landau-type
inequality is saturated for all (boundary and non-boundary) correlations with
unbiased marginals. We conjecture that the inequality is saturated for every
extremal correlation even with biased marginals, i.e., most of nonlocal
correlations in
the simplest Bell scenario obey a simple and unified rule. Second, the maximum
limit of one message bit set by the information causality principle splits into
the two trace distancelike quantities, which must not exceed 1 and
individually set a limit to determine the quantum boundaries. Namely, the
maximalness of the orthogonality (or vanishment of the above mentioned
randomness) play an important role in determining some of the quantum
boundaries.

%%%%%%%%%%%%%%%%%%%%%%%%%%%%%%%%%%%%%%%%%%%%%%%%%%%%%%%%%%%%%%%%%%%%%%%%%%%%%%%
%
%%%%%%%%%%%%%%%%%%%%%%%%%%%%%%%%%%%%%%%%%%%%%%%%%%%%%%%%%%%%%%%%%%%%%%%%%%%%%%%
\begin{acknowledgments}
This work was supported by JSPS KAKENHI Grant No. 24540405. 
\end{acknowledgments}

\appendix

%%%%%%%%%%%%%%%%%%%%%%%%%%%%%%%%%%%%%%%%%%%%%%%%%%%%%%%%%%%%%%%%%%%%%%%%%%%%%%%
%
%%%%%%%%%%%%%%%%%%%%%%%%%%%%%%%%%%%%%%%%%%%%%%%%%%%%%%%%%%%%%%%%%%%%%%%%%%%%%%%
\section{Properties of $\tilde D$}
\label{sec: Properties of D}

Some properties of the trace distancelike quantity
%%%%%%%%%%%%%%%%%%%%%%%%%%%%%%%%%%%%%%%%%%%%%%%%%%%%%%%%%%%%%%%%%%%%%%%%%%%%%%%
\begin{equation}
\tilde D(\rho,\sigma)=\max_{X}\frac{\hbox{tr}X(\rho-\sigma)}
{\sqrt{\hbox{tr}X^2(\rho+\sigma)}},
\label{eq: definition of tilde D}
\end{equation}
%%%%%%%%%%%%%%%%%%%%%%%%%%%%%%%%%%%%%%%%%%%%%%%%%%%%%%%%%%%%%%%%%%%%%%%%%%%%%%%
between subnormalized $\rho$ and $\sigma$ are proved here.
Without loss of generality, we can assume $\hbox{tr}(\rho+\sigma)\!=\!1$,
otherwise renormalize $\rho$ and $\sigma$.
Since the maximization is taken over all Hermitian operators $X$,
the constraint of $\hbox{tr}X^2(\rho+\sigma)\!=\!1$ does not alter the
optimization result, and hence let us maximize
%%%%%%%%%%%%%%%%%%%%%%%%%%%%%%%%%%%%%%%%%%%%%%%%%%%%%%%%%%%%%%%%%%%%%%%%%%%%%%%
\begin{equation}
\hbox{tr}X(\rho-\sigma)-l(\hbox{tr}X^2(\rho+\sigma)-1),
\end{equation}
%%%%%%%%%%%%%%%%%%%%%%%%%%%%%%%%%%%%%%%%%%%%%%%%%%%%%%%%%%%%%%%%%%%%%%%%%%%%%%%
where $l$ is the Lagrange multiplier. The extremal condition with respect to
the small deviation of $X\!\rightarrow\!X+\Delta X$, where $\Delta X$ is any
Hermitian operator, is given by
%%%%%%%%%%%%%%%%%%%%%%%%%%%%%%%%%%%%%%%%%%%%%%%%%%%%%%%%%%%%%%%%%%%%%%%%%%%%%%%
\begin{equation}
Y\equiv\rho-\sigma-l(\rho+\sigma)X-lX(\rho+\sigma)=0.
\label{eq: Extremal condition 1}
\end{equation}
%%%%%%%%%%%%%%%%%%%%%%%%%%%%%%%%%%%%%%%%%%%%%%%%%%%%%%%%%%%%%%%%%%%%%%%%%%%%%%%
In the case of pure states, let $\rho\!=\!p|0\rangle\langle0|$ and
$\sigma\!=\!(1-p)|\phi\rangle\langle\phi|$ where
$|\phi\rangle=\cos\phi|0\rangle+\sin\phi|1\rangle$, and let
$X=\left(\begin{array}{cc} a & b \cr b & c \end{array}\right)$.
From $\langle 1|Y|1\rangle\!=\!0$ and $\langle\phi_\perp|Y|\phi_\perp\rangle\!=\!0$, where $|\phi_\perp\rangle=\sin\phi|0\rangle-\cos\phi|1\rangle$, we have
$a+c=0$ and hence $X^2\!=\!(a^2+b^2)I=I$ so that 
$\hbox{tr}X^2(\rho+\sigma)\!=\!1$.
Since 
%%%%%%%%%%%%%%%%%%%%%%%%%%%%%%%%%%%%%%%%%%%%%%%%%%%%%%%%%%%%%%%%%%%%%%%%%%%%%%%
\begin{equation}
\bar D(\rho,\sigma)\equiv
\hbox{tr}|\rho-\sigma|=\max_{X^2=I}\hbox{tr}X(\rho-\sigma),
\end{equation}
%%%%%%%%%%%%%%%%%%%%%%%%%%%%%%%%%%%%%%%%%%%%%%%%%%%%%%%%%%%%%%%%%%%%%%%%%%%%%%%
we have $\tilde D(\rho,\sigma)\!=\!\bar D(\rho,\sigma)$ in the case of pure
states. In the other general cases,
$\tilde D(\rho,\sigma)\!\ge\!\bar D(\rho,\sigma)$ is obvious.

Moreover, let $\rho\!-\!\sigma\!=\!Q\!-\!S$ where $Q$ and $S$ are positive
operators with orthogonal support. Since
$|\rho\!-\!\sigma|\!\le\!\rho\!+\!\sigma$,
$\hbox{tr}(Q\!+\!S)\!=\!\hbox{tr}|\rho\!-\!\sigma|$,
and $\tilde D(\rho,\sigma)\!=\!1$ if $\rho$ and $\sigma$ are orthogonal,
we have
%%%%%%%%%%%%%%%%%%%%%%%%%%%%%%%%%%%%%%%%%%%%%%%%%%%%%%%%%%%%%%%%%%%%%%%%%%%%%%%
\begin{eqnarray}
\tilde D(\rho,\sigma)
&\le&\max_{X}\frac{\hbox{tr}X(\rho-\sigma)}{\sqrt{\hbox{tr}X^2|\rho-\sigma|}}
=\max_{X}\frac{\hbox{tr}X(Q-S)}{\sqrt{\hbox{tr}X^2(Q+S)}} \cr
&=&\sqrt{\hbox{tr}(Q+S)}=\sqrt{\bar D(\rho,\sigma)}.
\end{eqnarray}
%%%%%%%%%%%%%%%%%%%%%%%%%%%%%%%%%%%%%%%%%%%%%%%%%%%%%%%%%%%%%%%%%%%%%%%%%%%%%%%
Therefore,
$\bar D(\rho,\sigma)\!\le\!\tilde D(\rho,\sigma)\!\le\!\sqrt{\bar D(\rho,\sigma)}$,
and hence $\tilde D(\rho,\sigma)\!=\!1$ if and only if
$\bar D(\rho,\sigma)\!=\!1$.

The optimization with respect to $X$ in Eq.\ (\ref{eq: definition of tilde D})
can be
performed analytically as follows. Let $|i\rangle$ be the eigenstate of
$\rho+\sigma$, i.e., $(\rho+\sigma)|i\rangle=\lambda_i|i\rangle$, and
$x_{ij}\equiv\langle i|X|j \rangle$. We then have
%%%%%%%%%%%%%%%%%%%%%%%%%%%%%%%%%%%%%%%%%%%%%%%%%%%%%%%%%%%%%%%%%%%%%%%%%%%%%%%
\begin{eqnarray}
\hbox{tr}X^2(\rho+\sigma)
&=&\sum_{i}\lambda_i(x_{ii})^2+ \sum_{j>i}(\lambda_i+\lambda_j)|x_{ij}|^2 \cr
&=&\sum_{i}(x'_{ii})^2+\sum_{j>i}|x'_{ij}|^2=1,
\label{eq: Constraint for tilde D}
\end{eqnarray}
%%%%%%%%%%%%%%%%%%%%%%%%%%%%%%%%%%%%%%%%%%%%%%%%%%%%%%%%%%%%%%%%%%%%%%%%%%%%%%%
where $x'_{ii}\equiv\sqrt{\lambda_i}x_{ii}$ and
$x'_{ij}\equiv\sqrt{\lambda_i\!+\!\lambda_j}x_{ij}$ for $i\!\ne\!j$. Moreover,
using $a_{ij}=\langle i|\rho-\sigma|j\rangle$, we have
%%%%%%%%%%%%%%%%%%%%%%%%%%%%%%%%%%%%%%%%%%%%%%%%%%%%%%%%%%%%%%%%%%%%%%%%%%%%%%%
\begin{eqnarray}
\hbox{tr}X(\rho-\sigma)&=&\sum_{i}x_{ii}a_{ii}
+ \sum_{j>i}(x_{ij}a^{*}_{ij}+x^{*}_{ij}a_{ij}) \cr
&=&\sum_{i}\frac{x'_{ii}a_{ii}}{\sqrt{\lambda_i}}
+\sum_{j>i}\frac{x'_{ij}a^{*}_{ij}+x'^{*}_{ij}a_{ij}}
{\sqrt{\lambda_i+\lambda_j}}.
\label{eq: Expression of tilde D}
\end{eqnarray}
%%%%%%%%%%%%%%%%%%%%%%%%%%%%%%%%%%%%%%%%%%%%%%%%%%%%%%%%%%%%%%%%%%%%%%%%%%%%%%%
Since $\tilde D(\rho,\sigma)$ is given by the maximum of
Eq.\ (\ref{eq: Expression of tilde D}) under the constraint of
Eq.\ (\ref{eq: Constraint for tilde D}), we have
%%%%%%%%%%%%%%%%%%%%%%%%%%%%%%%%%%%%%%%%%%%%%%%%%%%%%%%%%%%%%%%%%%%%%%%%%%%%%%%
\begin{equation*}
\tilde D(\rho,\sigma)=
\big(\sum_{i}\frac{(a_{ii})^{2}}{\lambda_i}+
\sum_{j>i}\frac{4|a_{ij}|^2}{\lambda_i+\lambda_j}\big)^{\frac{1}{2}}
=\big(\sum_{ij}\frac{2|a_{ij}|^2}{\lambda_i+\lambda_j}\big)^{\frac{1}{2}}.
\end{equation*}
%%%%%%%%%%%%%%%%%%%%%%%%%%%%%%%%%%%%%%%%%%%%%%%%%%%%%%%%%%%%%%%%%%%%%%%%%%%%%%%

%%%%%%%%%%%%%%%%%%%%%%%%%%%%%%%%%%%%%%%%%%%%%%%%%%%%%%%%%%%%%%%%%%%%%%%%%%%%%%%
%
%%%%%%%%%%%%%%%%%%%%%%%%%%%%%%%%%%%%%%%%%%%%%%%%%%%%%%%%%%%%%%%%%%%%%%%%%%%%%%%
\section{Landau-type inequality}
\label{sec: Landau-type inequality}

In the CHSH expression with general weight parameters,
if we define $X_{x}\!=\!u_{x0}B_0\!+\!(-1)^x u_{x1} B_1$, we have
$X^{2}_{0}\!+\!X^{2}_{1}\!=\!\sum_{xy}u^{2}_{xy}I$ by virtue of
$u_{00}u_{01}\!=\!u_{10}u_{11}$, and hence obtain
Eq.\ (\ref{eq: weighted CHSH bound}).
When $\tilde D_0,\tilde D_1\!>\!0$, we have
%%%%%%%%%%%%%%%%%%%%%%%%%%%%%%%%%%%%%%%%%%%%%%%%%%%%%%%%%%%%%%%%%%%%%%%%%%%%%%%
\begin{eqnarray}
\sum_{x}\big(u_{x0}\tilde C_{x0}+(-1)^{x}u_{x1}\tilde C_{x1}\big)^2\le\sum_{xy}u^{2}_{xy}.
\end{eqnarray}
%%%%%%%%%%%%%%%%%%%%%%%%%%%%%%%%%%%%%%%%%%%%%%%%%%%%%%%%%%%%%%%%%%%%%%%%%%%%%%%
Noticing $\tilde C^{2}_{xy}\!\le\!1$ and using
$v_{xy}\!\equiv\!u_{xy}(1\!-\!\tilde C^{2}_{xy})^{\frac{1}{2}}$, this is
rewritten as
%%%%%%%%%%%%%%%%%%%%%%%%%%%%%%%%%%%%%%%%%%%%%%%%%%%%%%%%%%%%%%%%%%%%%%%%%%%%%%%
\begin{equation*}
(\tilde C_{00} \tilde C_{01}\!-\!\tilde C_{10}\tilde C_{11})v_{00}v_{01}
\le
(1\!-\!\tilde C^{2}_{00})^{\frac{1}{2}}(1\!-\!\tilde C^{2}_{01})^{\frac{1}{2}}
\sum_{xy}{\textstyle \frac{1}{2}}v^{2}_{xy}.
\end{equation*}
%%%%%%%%%%%%%%%%%%%%%%%%%%%%%%%%%%%%%%%%%%%%%%%%%%%%%%%%%%%%%%%%%%%%%%%%%%%%%%%
Since $u_{00}u_{01}\!=\!u_{10}u_{11}$, it can be seen that
%%%%%%%%%%%%%%%%%%%%%%%%%%%%%%%%%%%%%%%%%%%%%%%%%%%%%%%%%%%%%%%%%%%%%%%%%%%%%%%
\begin{eqnarray*}
\lefteqn{(1\!-\!\tilde C^{2}_{00})^{\frac{1}{2}}(1\!-\!\tilde C^{2}_{01})^{\frac{1}{2}}
\sum_{xy}{\textstyle \frac{1}{2}}v^{2}_{xy}} \cr
&\!\!\ge\!\!&
(1\!-\!\tilde C^{2}_{00})^{\frac{1}{2}}(1\!-\!\tilde C^{2}_{01})^{\frac{1}{2}}
(|v_{00}v_{01}|+|v_{10}v_{11}|) \\
&\!\!=\!\!&
[(1\!-\!\tilde C^{2}_{00})^{\frac{1}{2}}(1\!-\!\tilde C^{2}_{01})^{\frac{1}{2}}
+(1\!-\!\tilde C^{2}_{10})^{\frac{1}{2}}(1\!-\!\tilde C^{2}_{11})^{\frac{1}{2}}]
|v_{00}v_{01}|,
\end{eqnarray*}
%%%%%%%%%%%%%%%%%%%%%%%%%%%%%%%%%%%%%%%%%%%%%%%%%%%%%%%%%%%%%%%%%%%%%%%%%%%%%%%
and we obtain Eq.\ (\ref{eq: extended Landau bound}). In the case of
$\tilde D_{x_0}\!=\!0$ for either $x_0\!=\!0$ or $1$,
$C_{x_0y}\!=\!0$ holds because $\tilde D_{x_0}\!\ge\!|C_{x_0y}|$ by definition,
and we have Eq.\ (\ref{eq: extended Landau bound}) in which
$\tilde C_{x_00}\!=\!\tilde C_{x_01}\!=\!0$.

Similarly, for a tilted CHSH expression, we have
%%%%%%%%%%%%%%%%%%%%%%%%%%%%%%%%%%%%%%%%%%%%%%%%%%%%%%%%%%%%%%%%%%%%%%%%%%%%%%%
\begin{eqnarray}
\lefteqn{
\sum_{xy}s_x u_{xy}(-1)^{xy}\langle(A_x\!+\!\epsilon_x I)\otimes B_y\rangle
} \cr
&\le&\big[\sum_{xy}u^{2}_{xy}\big]^{\frac{1}{2}}
\big[\sum_x s^{2}_x(\tilde D^{\epsilon}_x)^2)\big]^{\frac{1}{2}},
\label{eq: weighted tilted CHSH bound}
\end{eqnarray}
%%%%%%%%%%%%%%%%%%%%%%%%%%%%%%%%%%%%%%%%%%%%%%%%%%%%%%%%%%%%%%%%%%%%%%%%%%%%%%%
where $u_{00}u_{01}\!=\!u_{10}u_{11}$ again, $\epsilon_x$ is real, and
%%%%%%%%%%%%%%%%%%%%%%%%%%%%%%%%%%%%%%%%%%%%%%%%%%%%%%%%%%%%%%%%%%%%%%%%%%%%%%%
\begin{equation}
\tilde D^{\epsilon}_{x}\!\equiv\!
\max_X \frac{\hbox{tr}X[(1+\epsilon_x)\rho_{0|x}-(1-\epsilon_x)\rho_{1|x}]}{\sqrt{\hbox{tr}X^2(\rho_{0|x}+\rho_{1|x})}}.
\end{equation}
%%%%%%%%%%%%%%%%%%%%%%%%%%%%%%%%%%%%%%%%%%%%%%%%%%%%%%%%%%%%%%%%%%%%%%%%%%%%%%%
In the same way as
Appendix \ref{sec: Properties of D}, it is not difficult to see
$(\tilde D^{\epsilon}_{x})^2\!=\!\tilde D_{x}^2\!+\!2\epsilon_x\langle A_x\rangle\!+\!\epsilon_{x}^2$.
When
$\tilde D^{2}_0\!>\!\langle A_0\rangle^{2}$ and
$\tilde D^{2}_1\!>\!\langle A_1\rangle^{2}$,
the necessary and sufficient
condition of Eq.\ (\ref{eq: weighted tilted CHSH bound}) is again
Eq.\ (\ref{eq: extended Landau bound}) but with
%%%%%%%%%%%%%%%%%%%%%%%%%%%%%%%%%%%%%%%%%%%%%%%%%%%%%%%%%%%%%%%%%%%%%%%%%%%%%%%
\begin{equation}
\tilde C_{xy}\equiv \frac{C_{xy}-\langle A_x\rangle\langle B_y\rangle}
{(\tilde D^{2}_x-\langle A_x\rangle^2)^{\frac{1}{2}}
(1-\langle B_y\rangle^2)^{\frac{1}{2}}}.
\label{eq: definition of G}
\end{equation}
%%%%%%%%%%%%%%%%%%%%%%%%%%%%%%%%%%%%%%%%%%%%%%%%%%%%%%%%%%%%%%%%%%%%%%%%%%%%%%%
This is an extension of the NPA inequality \cite{Navascues07a,Navascues08a}.

%%%%%%%%%%%%%%%%%%%%%%%%%%%%%%%%%%%%%%%%%%%%%%%%%%%%%%%%%%%%%%%%%%%%%%%%%%%%%%%
%
%%%%%%%%%%%%%%%%%%%%%%%%%%%%%%%%%%%%%%%%%%%%%%%%%%%%%%%%%%%%%%%%%%%%%%%%%%%%%%%
\section{Extremal correlation}
\label{sec: Extremal correlation}

Under the parametrization of Eq.\ (\ref{eq: Parametrization 1}), the
expectation of the Bell expression
$\sum_{abxy} V_{abxy}p(ab|xy)$ with
$p(ab|xy)\!=\!\langle\psi| P_{a|x}\!\otimes\!Q_{b|y}|\psi\rangle$ is
maximized when $\rho\!=\!|\psi\rangle\langle\psi|$ is real symmetric.
A necessary and sufficient condition for the saturation of
Eq.\ (\ref{eq: extended Landau bound}) is that it is possible to assign
nontrivial values to $u_{xy}$ such that Eq.\ (\ref{eq: weighted CHSH bound}) is
saturated (and $u_{00}u_{01}\!=\!u_{10}u_{11}$). This is possible only when
$X_x\!=\!\sum_y (-1)^{xy}u_{xy}B_y$ agrees with the operator of maximizing
$\tilde D_x$. Note that $\rho$ is pure and the projector is rank 1,
$\rho_{a|x}$ is also pure and $\tilde D_x\!=\!\bar D_x$. Since the operator of
maximizing $\tilde D_x$ is then unique up to the normalization, we have
$X_x\!\propto\!\cos\phi^{B}_x\sigma_1\!+\!\sin\phi^{B}_x\sigma_3$ hence
$\sum_y(-1)^{xy}u_{xy}\sin(\phi^{B}_x\!-\!\theta^{B}_y)\!=\!0$.
In order that $u_{00}u_{01}\!=\!u_{10}u_{11}$,
$-\sin(\phi^{B}_1\!-\!\theta^{B}_0)\sin(\phi^{B}_0\!-\!\theta^{B}_0)u^{2}_{00}
\!=\!\sin(\phi^{B}_0\!-\!\theta^{B}_1)\sin(\phi^{B}_1\!-\!\theta^{B}_1)u^{2}_{11}$ and
$-\sin(\phi^{B}_0\!-\!\theta^{B}_1)\sin(\phi^{B}_1\!-\!\theta^{B}_1)u^{2}_{01}
\!=\!\sin(\phi^{B}_1\!-\!\theta^{B}_0)\sin(\phi^{B}_0\!-\!\theta^{B}_0)u^{2}_{10}$
must hold, and we obtain Eq.\ (\ref{eq: Saturation condition B}).
Since there are no other constraints for $u_{xy}$, we can assign nontrivial
values to $u_{xy}$ if Eq.\ (\ref{eq: Saturation condition B}) is satisfied.

%%%%%%%%%%%%%%%%%%%%%%%%%%%%%%%%%%%%%%%%%%%%%%%%%%%%%%%%%%%%%%%%%%%%%%%%%%%%%%%
%
%%%%%%%%%%%%%%%%%%%%%%%%%%%%%%%%%%%%%%%%%%%%%%%%%%%%%%%%%%%%%%%%%%%%%%%%%%%%%%%
\section{Evaluation of $1-H(a_p|\vec x\rho_B)$}
\label{sec: Evaluation of 1-H}

Let us denote Bob's guess for the parity bit $a_p$ (for a given
$\vec x$) under the individual measurement strategy for boxes by $b_p$, the
conditional probability by $P_{a_p|b_p}$, and the other probabilities
similarly. Let us then evaluate the leading term of $H(a_p|b_p)$ given by 
%%%%%%%%%%%%%%%%%%%%%%%%%%%%%%%%%%%%%%%%%%%%%%%%%%%%%%%%%%%%%%%%%%%%%%%%%%%%%%%
\begin{eqnarray*}
\lefteqn{H(a_p|b_p)=P_{b_p=0}h(P_{0|0})+P_{b_p=1}h(P_{1|1})} \cr
&\!\!\!\approx\!\!\!&1-\frac{1}{2\ln2}\big[
P_{b_p=0}(2P_{0|0}-1)^2+P_{b_p=1}(2P_{1|1}-1)^2\big] \cr
&\!\!\!=\!\!\!&1-\frac{1}{2\ln2}\big[
\frac{(P_{a_p=b_p}-P_{a_p=1})^2}{P_{b_p=0}}
+\frac{(P_{a_p=b_p}-P_{a_p=0})^2}{P_{b_p=1}}\big],
\end{eqnarray*}
%%%%%%%%%%%%%%%%%%%%%%%%%%%%%%%%%%%%%%%%%%%%%%%%%%%%%%%%%%%%%%%%%%%%%%%%%%%%%%%
for $k\!\gg\!1$ and $n\!-\!k\!\gg\!1$ with $k$ being the number of 0 in a
given $\vec x$ (see also \cite{Bennett96d}). Since the optimal success
probability of guessing $a$ for a single box is $(1\!+\!\bar D_x)/2$, the
optimal probability for $a_p$ is given by
$P_{a_p=b_p}\!=\!(1+\bar D^{k}_{0} \bar D^{n-k}_{1})/2$.
Using Alice's marginals $p(a|x)$ of a single box, we have
%%%%%%%%%%%%%%%%%%%%%%%%%%%%%%%%%%%%%%%%%%%%%%%%%%%%%%%%%%%%%%%%%%%%%%%%%%%%%%%
\begin{equation}
P_{a_p=0}=\frac{1}{2}\big[1+\left(2p(0|0)-1\right)^k\left(2p(0|1)-1\right)^{n-k}\big].
\end{equation}
%%%%%%%%%%%%%%%%%%%%%%%%%%%%%%%%%%%%%%%%%%%%%%%%%%%%%%%%%%%%%%%%%%%%%%%%%%%%%%%
Suppose that $\rho_{0|x}$ and $\rho_{1|x}$ are nonorthogonal.
Moreover, $\hbox{sup}(\rho_{0|x})$ and $\hbox{sup}(\rho_{1|x})$
are not identical in general. This implies
$|2p(0|x)\!-\!1|\!<\!\bar D_x\!<\!1$, and hence the leading term comes from
$\bar D^{k}_0 \bar D^{n-k}_1$. As a result, since
$H(a_p|b_p)\approx 1-\bar D^{2k}_{0} \bar D^{2(n-k)}_{1}/(2\ln2)$,
we have
%%%%%%%%%%%%%%%%%%%%%%%%%%%%%%%%%%%%%%%%%%%%%%%%%%%%%%%%%%%%%%%%%%%%%%%%%%%%%%%
\begin{eqnarray*}
\lefteqn{1-H(a_p|\vec x\rho_B)=1-\frac{1}{2^n}\sum_{\vec x}H(a_p|b_p)} \cr
&\!\!\!\approx\!\!\!&
\frac{1}{2\ln2} \frac{1}{2^n} \sum_{k=0}^{n} \binom{n}{k}
\bar D^{2k}_0 \bar D^{2(n-k)}_1
=\frac{1}{2\ln2}\big(\frac{\bar D^{2}_0+\bar D^{2}_1}{2}\big)^n.
\end{eqnarray*}
%%%%%%%%%%%%%%%%%%%%%%%%%%%%%%%%%%%%%%%%%%%%%%%%%%%%%%%%%%%%%%%%%%%%%%%%%%%%%%%

%Chap. 15, pp. 335-366.
%%%%%%%%%%%%%%%%%%%%%%%%%%%%%%%%%%%%%%%%%%%%%%%%%%%%%%%%%%%%%%%%%%%%%%%%%%%%%%%
%\bibliography{personal}
%\bibliographystyle{apsrev}
%%%%%%%%%%%%%%%%%%%%%%%%%%%%%%%%%%%%%%%%%%%%%%%%%%%%%%%%%%%%%%%%%%%%%%%%%%%%%%%

%%%%%%%%%%%%%%%%%%%%%%%%%%%%%%%%%%%%%%%%%%%%%%%%%%%%%%%%%%%%%%%%%%%%%%%%%%%%%%%
%
%%%%%%%%%%%%%%%%%%%%%%%%%%%%%%%%%%%%%%%%%%%%%%%%%%%%%%%%%%%%%%%%%%%%%%%%%%%%%%%

\end{document}